\documentclass[preprint,showpacs,preprintnumbers,superscriptaddress,amsmath,amssymb]{revtex4}

\usepackage{graphicx}
\usepackage{dcolumn}
\usepackage{bm}
\newcommand{\astra}{{\sc Astra }}

\newcommand{\mean}[1]{\mbox{$\langle{#1}\rangle$}}

\begin{document}

\title{Generation of angular-momentum-dominated  \\
 electron beams from a photoinjector}

\affiliation{University of Chicago, Chicago, IL 60637, USA}
\affiliation{Fermi National Accelerator Laboratory, Batavia, IL
60510, USA} \affiliation{Argonne National Laboratory, Argonne, IL
60439, USA} \affiliation{Northern Illinois University, DeKalb, IL
60115, USA} \affiliation{Lawrence Berkeley National Laboratory,
Berkeley, CA 94720, USA}
\affiliation{University of Rochester,
Rochester, NY 14627, USA}
\author{Y.-E Sun}
\email[Corresponding author. Electronic address:
]{yinesun@uchicago.edu} \affiliation{University of Chicago, Chicago,
IL 60637, USA}
\author{P. Piot}
\email[Corresponding author. Electronic address: ]{piot@fnal.gov}
\affiliation{Fermi National Accelerator Laboratory, Batavia, IL
60510, USA}
\author{\\K.-J. Kim}
\altaffiliation[Also at ]{University of Chicago, Chicago, IL 60637, USA}
\affiliation{Argonne National Laboratory, Argonne, IL 60439, USA}
\author{N. Barov}
\altaffiliation[Now at ]{Far-tech Inc, San Diego, CA 92121, USA}
\affiliation{Northern Illinois University, DeKalb, IL 60115, USA}
\author{S. Lidia}
\affiliation{Lawrence Berkeley National Laboratory, Berkeley, CA
94720, USA}
\author{J. Santucci}
\affiliation{Fermi National Accelerator Laboratory, Batavia, IL
60510, USA}
\author{R. Tikhoplav}
\affiliation{University of Rochester, Rochester, NY 14627, USA}
\author{J. Wennerberg}
\altaffiliation[Now at ]{Purdue University, West Lafayette, IN
47907, USA} \affiliation{Fermi National Accelerator Laboratory,
Batavia, IL 60510, USA}

\date{\today}

\begin{abstract}
Various projects under study require an angular-momentum-dominated
electron beam generated by a photoinjector. Some of the proposals
directly use the angular-momentum-dominated beams (e.g. electron
cooling of heavy ions), while others require the beam to be
transformed into a flat beam (e.g. possible electron injectors for
light sources and linear colliders). In this paper, we report our
experimental study of an angular-momentum-dominated beam produced in
a photoinjector, addressing the dependencies of angular momentum on
initial conditions. We also briefly discuss the removal of angular
momentum. The results of the experiment, carried out at the
Fermilab/NICADD Photoinjector Laboratory, are found to be in good
agreement with theoretical and numerical models.
\end{abstract}
\maketitle
\section{Introduction}
Angular-momentum-dominated electron beams generated by
photoinjectors have direct applications in several accelerator
proposals presently under consideration, either in the field of
high-energy colliders or accelerator-based light sources. In
Reference~\cite{benzvi}, an angular-momentum-dominated, or
``magnetized", beam is proposed to be accelerated to $\sim 50$~MeV
and used for electron beam cooling~\cite{budker,derbenevmag} of
ion beams in the relativistic heavy ion collider (RHIC). In such a
scheme, the electron beam propagates together with the ion beam at
the same velocity. Collisions of ions with electrons lead to a
transfer of thermal motion from the ion to the electron beam. As
the two beams co-propagate, the electron-ion effective interaction
length is increased due to the helical trajectory of the electron
in the magnetic field, thereby improving the cooling efficiency.
The cooling rate is then mainly determined by the longitudinal
momentum spread of the electron beam, which can be made much
smaller than the transverse one. Reference~\cite{brinkmann}
concerns the photoinjector production of flat beams, i.e. a beam
with high transverse emittance ratio. The technique consists of
manipulating an angular-momentum-dominated beam produced by a
photoinjector using the linear transformation described in
Reference~\cite{derbenev}. The latter linear transformation
removes the angular momentum and results in a flat beam. In the
context of linear collider proposals, where a flat beam at the
interaction point is needed to reduce beamstrahlung~\cite{yokoya},
the development of a flat-beam electron source is an attractive
idea since it could simplify or eliminate the need for an electron
damping ring. The flat beam technique is also proposed for
generation of ultrashort X-ray pulses by making use of the smaller
dimension of the flat beam ~\cite{lux}, and also in enhancing
beam-surface interaction in a Smith-Purcell radiator~\cite{kjk2}
or in an image charge undulator~\cite{smithpurcell}. A
proof-of-principle experiment conducted at the Fermilab/NICADD
Photoinjector Laboratory (FNPL)\footnote{NICADD is an acronym for
Northern Illinois Center for Accelerator and Detector
Development.} has demonstrated the flat beam
production~\cite{edwards,edwardspac01}, where an emittance ratio
of $50$ was reported.

In this paper we report on recent results pertaining to the
experimental investigation of some properties of an
angular-momentum-dominated beam. We also briefly address the removal
of angular momentum and the subsequent generation of a flat beam.
Producing flat beams is our primary motivation for the present
studies.

In Section~\ref{sec:THEORY} we briefly summarize theoretical aspects
of the photoinjector production of angular-momentum-dominated beams.
In Section~\ref{sec:EXP} we describe the experimental set-up of
FNPL. Sections~\ref{sec:MOM} and~\ref{sec:REMOVAL} are dedicated to
experimental results and their comparisons to theory and numerical
simulations. Our conclusions appear in Section~\ref{sec:FUTURE}.
\section{theoretical background}~\label{sec:THEORY}
In this section we assume the beam and external focusing forces to
be cylindrically symmetric. The cylindrical symmetry implies the
conservation of the canonical angular momentum of each electron. In
an axial magneto static field $B_z(z)$, the canonical angular
momentum of an electron, $L$, in circular cylindrical coordinates
$(r, \phi, z)$ is~\cite{Reiser}

\begin{eqnarray} \label{e:totalcam}
L = \gamma m r^2 \dot{\phi}+ \frac{1}{2}e B_z(z) r^2,
\end{eqnarray}
where $\gamma$ is the Lorentz factor, $\dot{\phi}$ the time
derivative of $\phi$, $m$ and $e$ are respectively the electron rest
mass and charge.

The average canonical angular momentum of the electrons, $\mean{L}$,
is obtained by averaging Eq.~(\ref{e:totalcam}) over the beam
distribution. At the photocathode location, we have
$\mean{\dot{\phi}}=0$ and
\begin{eqnarray} \label{e:camdef}
\mean{L}= \frac{1}{2}e B_0 \mean{r^2}= e B_0  \sigma_c^2,
\end{eqnarray}
where $\sigma_c=\sqrt{\mean{r^2}/2}$ is the transverse
root-mean-square (rms) beam size on the photocathode, $B_0=
B_z(z=0)$ is the axial magnetic field on the photocathode.

Outside the solenoidal field region, where $B_z$ vanishes, an
electron acquires mechanical angular momentum due to the torque
exerted on it in the transition region. Since $B_z(z)=0$, the second
term of Eq.~(\ref{e:totalcam}) vanishes and the canonical angular
momentum is given by the first term of  Eq.~(\ref{e:totalcam}),
which is the mechanical angular momentum. It is convenient to
normalize $\mean{L}$ with the axial momentum $p_z$, and introduce
the quantity ${\cal L}$ given by
\begin{equation}
{\mathcal L}=\frac{\mean{L}}{2p_z}= \kappa \sigma_c^2,
\end{equation}
where $\kappa = eB_0/(2p_z)$.

The beam angular momentum can be removed by means of a properly
designed skew quadrupole section~\cite{burov, brinkmann2, BND-PRE}
and the beam is transformed into a flat beam (see
section~\ref{sec:REMOVAL}). The flat beam transverse emittances
after the skew quadrupole section, $\epsilon_{\pm}$, are given
by~\cite{BND-PRE,kjk}:
\begin{eqnarray} \label{eq:gen}
\epsilon_{\pm}= \sqrt{\epsilon_u^2 + {\cal L}^2} \pm {\cal L}.
\end{eqnarray}
Here $\epsilon_u$ is the uncorrelated transverse emittance prior to
the skew quadrupole section. Note that the four dimensional
emittance is conserved since $\epsilon_u^2=\epsilon_+\epsilon_-$.

The evolution of the transverse rms beam size of a relativistic
electron bunch in a drift is given by the envelope
equation~\cite{reiserenv}
\begin{eqnarray} \label{eq:envelope}
\sigma'' - \frac{K}{4\sigma} -\frac{\epsilon_u^2}{\sigma^3} -\frac{
{\cal L}^2}{\sigma^3} =0,
\end{eqnarray}
where $\sigma$ is the transverse rms size, $K = \frac{2I}{I_0
\gamma^3}$ is the generalized perveance, $I$ is the absolute value
of the instantaneous beam current and $I_0$ is the Alfv\'en current
for electrons ($\sim 17$ kA). The second, third and fourth terms
respectively represent the effects due to space charge, emittance
and the angular momentum. For low energy beam, the space charge term
is important. However, for the typical operating conditions
considered in this paper, e.g., $\gamma \approx 30$, bunch charge
$\approx 0.5$~nC, rms beam duration $\sigma_t \approx 4$~ps, $\sigma
\approx 1.25$~mm~\cite{yesBD1}, $\gamma \epsilon_u \approx 4$ mm
mrad~\cite{lidia}, $\gamma \mathcal{L} \approx 20$~mm mrad, the
fourth term Eq.~\ref{eq:envelope} is much greater than the second
and the third term. Such a beam is said to be angular momentum
dominated.

\section{experimental setup}~\label{sec:EXP}
\begin{figure*}[hbt]
\includegraphics*[angle=0,width=150mm]{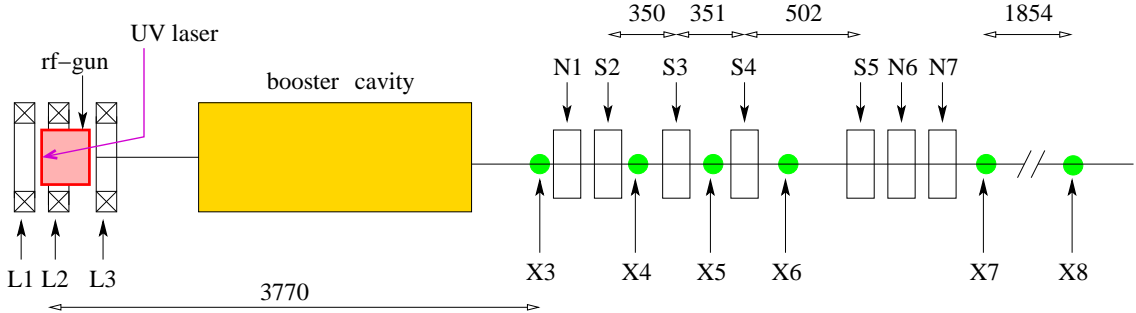}
\caption{Overview of the FNPL beamline. Here only the elements
pertaining to the flat-beam experiment are shown. The letters
represents solenoidal magnetic lenses (L), normal (N) and skew (S)
quadrupoles, and diagnostic stations (X). Dimensions are in mm.
\label{fig:injector}}
\end{figure*}
The experimental production and characterization of
angular-momentum-dominated electron beams were carried out at
FNPL.

The photoinjector incorporates a photoemission source consisting of
a $1+\frac{1}{2}$ cell cavity operating at 1.3~GHz, the so-called
radio frequency (rf) gun. An ultraviolet (UV) laser impinges a
cesium telluride photocathode located on the back plate of the rf
gun half cell. The thereby photoemitted electron bunch exits from
the rf-gun at 4 MeV/c and is immediately injected into a TESLA-type
superconducting cavity~\cite{teslacav} (henceforth referred to as
the booster cavity). The bunch momentum downstream of the booster
cavity is approximately 16~MeV/c when the cavity is operated to
yield the maximum energy gain. The typical operating conditions of
the main subsystems of the photoinjector are gathered in
Table~\ref{tab:injector}, and a block diagram of the facility is
depicted in Fig.~\ref{fig:injector}.

The transverse size of the UV drive-laser at the photocathode is set
by a remotely controllable iris. The laser temporal profile is a
Gaussian distribution with rms duration of $\sim$3.5~ps.

The rf gun is surrounded by three solenoidal magnetic lenses
independently powered.  This allows proper focusing of the electron
bunch while maintaining the desired magnetic field on the
photocathode.

Downstream of the booster cavity, the beamline includes a
round-to-flat-beam (RTFB) transformer, consisting of four skew
quadrupoles, that can be used to remove the mechanical angular
momentum.

Several optical transition radiation (OTR) or fluorescent
(YaG-based) screens serve as diagnostics to measure the beam's
transverse density at various locations in the beamline. Transverse
emittances can also be measured based on the
multislit~\cite{Lejeune, PP}, or quadrupole scan
techniques~\cite{Wiedemann}. The multislit mask used for emittance
measurements consists of a 6-mm-thick tungsten mask with
48~$\mu$m-wide slits spaced 1~mm apart.
\begin{table}[h!]
\begin{center}
\begin{tabular}{l c c}
\hline \hline parameter                       &      value       &
units  \\ \hline
laser injection phase           &  25   $\pm$ 5    & rf-deg \\
laser radius on cathode         &  [0.6, 1.6] $\pm$ 0.05  & mm     \\
laser pulse duration            &  3.5 $\pm$ 0.5  & ps     \\
bunch charge                    &  [0.2, 1.6]    & nC \\
$E_z$ on cathode                &  35  $\pm$ 0.2  & MV/m     \\
$B_0$ on cathode                &  [200, 1000]   & Gauss     \\
booster cavity acc. gradient    &  $\sim$ 12    & MV/m     \\
\hline \hline
\end{tabular}
\caption{\label{tab:injector} Typical settings for the photocathode
drive laser, rf gun, and accelerating section. Values in square
brackets correspond to the range used in the measurements. }
\end{center}
\end{table}
\section{Measurements of Canonical Angular Momentum}~\label{sec:MOM}
We now turn to the basic properties of the canonical angular
momentum. We especially investigate the conversion of the canonical
angular momentum of the photo-emitted electron bunch into mechanical
angular momentum downstream of the booster cavity.

\begin{figure}[hbt]
\includegraphics[width=.4\textwidth]{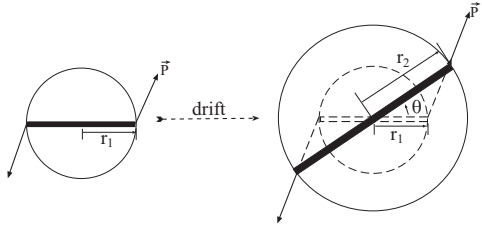}
\caption{Beam with angular-momentum-induced shearing while drifting.
The dark narrow rectangle represents a slit inserted into the
beamline to measure the shearing angle (see text for more details).
\label{fig:camdrif}}
\end{figure}
\begin{figure}[t]\centering
\includegraphics[width=.5\textwidth]{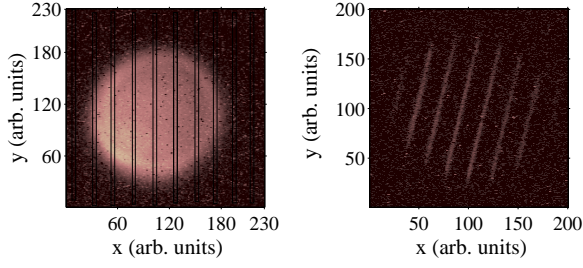}
\caption{Example of data set used for mechanical angular momentum
measurement. Beam transverse density on X3 (left) and observed
beamlets on X6 when the vertical multislit mask is inserted at X3
(right). The vertical lines superimposed on the X3 image is an
illustration of vertical slits when the multislit mask is
inserted.\label{fig:spotslit}}
\end{figure}
\begin{figure}[hbt] \centering
\includegraphics[width=0.48\textwidth]{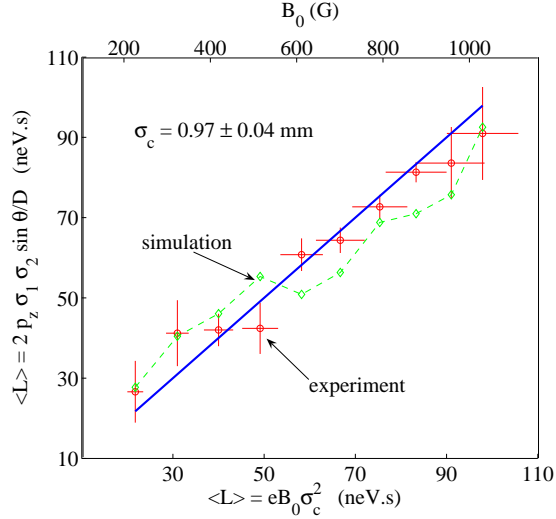}
\caption{Mechanical angular momentum from Eq.~(\ref{e:mechLsigma})
versus the canonical angular momentum calculated from
Eq.~(\ref{e:camdef}). The labels ``experiment" and ``simulation"
correspond respectively to experimentally measured data points and
simulated values found by modeling of the measurement technique. The
solid diagonal line is drawn simply to aid the eye.
\label{fig:LvsLb}}
\end{figure}
\begin{figure}[hbt] \centering
\includegraphics[width=0.48\textwidth]{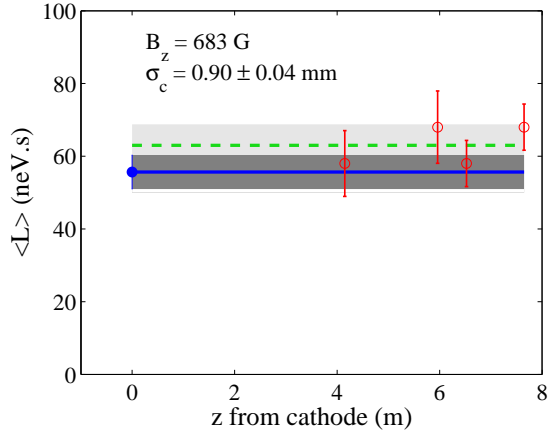}
\caption{Evolution of canonical angular momentum along the beamline.
At photocathode location (dot), canonical angular momentum is
calculated from Eq.~(\ref{e:camdef}) and solid line is this value
extended along z. At at other locations (circles), mechanical
angular momentum is obtained from Eq.~(\ref{e:mechLsigma}) and the
dashed line is the average. The shaded area covers the uncertainties
in the measurements either from Eq.~(\ref{e:camdef}) (darker strip)
or from Eq.~(\ref{e:mechLsigma}) (lighter strip).\label{fig:LvsZ}}
\end{figure}
\begin{figure}[hbt] \centering
\includegraphics[width=0.48\textwidth]{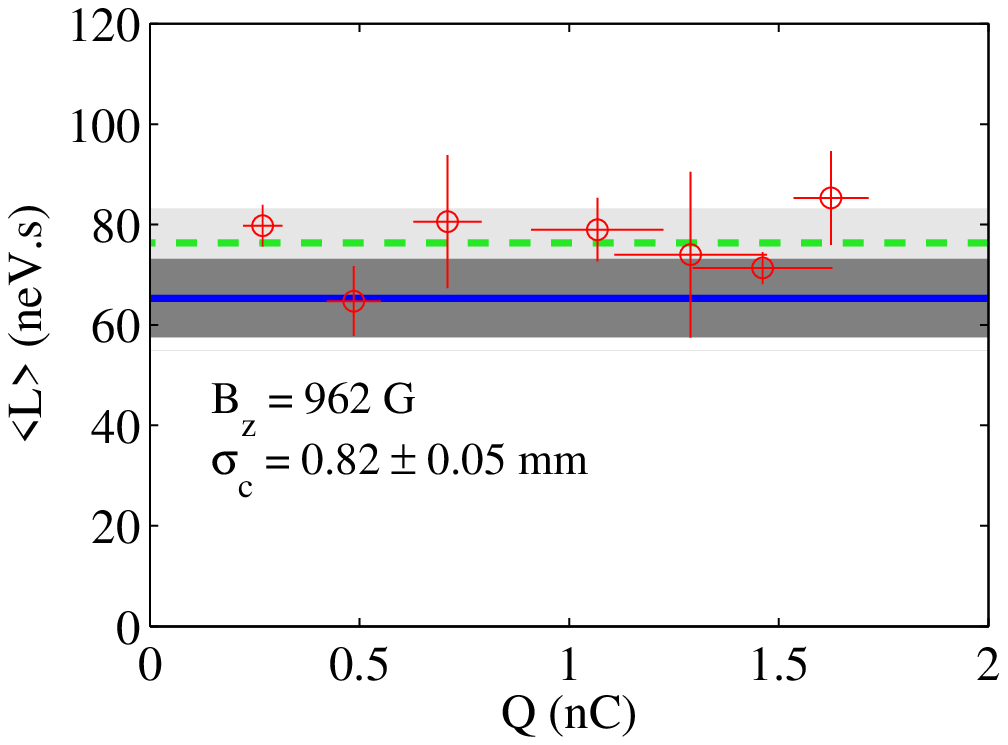}
\includegraphics[width=0.48\textwidth]{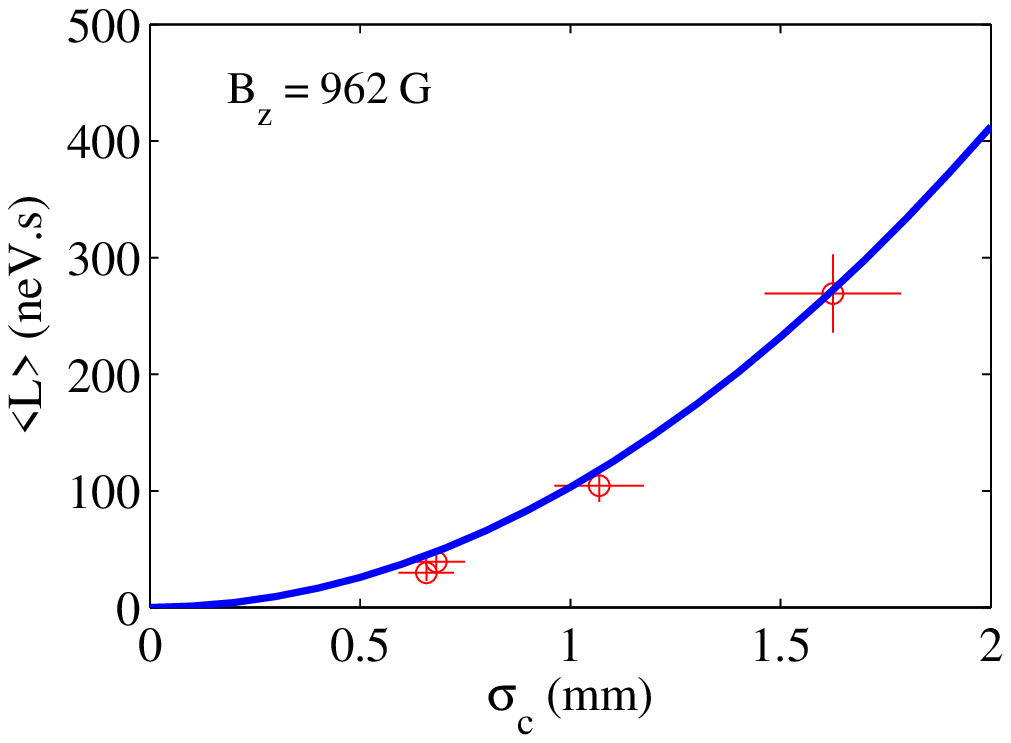}
\caption{Canonical angular momentum versus charge (top)  and
photocathode drive-laser spot size (bottom). The measured mechanical
angular momenta (circles) are compared with the theoretical value of
the canonical angular momentum calculated from the axial magnetic
field (solid line). In the top figure, the dashed line represents
the average value of all the data points, and the shaded area has
the same meaning as in Fig.~\ref{fig:LvsZ}. \label{fig:LvsQ}}
\end{figure}

The canonical angular momentum at the photocathode surface is
obtained from Eq.~(\ref{e:camdef}). Given the experimental settings
of the solenoidal lens currents, the magnetic field, $B_0$, is
inferred via simulations using the {\sc Poisson}~\cite{lanl}
program, which is bench-marked against calibration of the solenoidal
lenses~\cite{JPCthesis}. The value of $\sigma_c$ used in
Eq.~(\ref{e:camdef}) is directly measured from an image of the UV
laser on a ``virtual photocathode"
. The virtual photocathode is a calibrated UV-sensitive screen,
located outside of the vacuum chamber, being a one-to-one optical
image of the photocathode.

To elaborate the method used to measure the mechanical angular
momentum downstream of the booster cavity, we consider an electron
in a magnetic-field-free region at longitudinal location $z_1$
with transverse radial vector
$\boldsymbol{r}_1=r_1\hat{\boldsymbol{e}}_x$
($\hat{\boldsymbol{e}}_{x}$ stands for the $x$-axis unit vector).
After propagating through a drift space, the electron reaches
$\boldsymbol{r}_2$ at location $z_2$. Let $\theta= \angle
(\boldsymbol{r}_1, \boldsymbol{r}_2)$ be the angle between the two
aforementioned radial vectors ($\theta$ is henceforth referred to
as ``shearing angle"; see Fig.~\ref{fig:camdrif}). The mechanical
angular momentum of the electron, $\boldsymbol{L}$, is given by:
\begin{eqnarray}\label{e:demomechL1}
\boldsymbol{L}=r_1\hat{\boldsymbol{e}}_x\times\boldsymbol{P}=r_1p_{y}\hat{\boldsymbol{e}}_x\times\hat{\boldsymbol{e}}_y.
\end{eqnarray}
By introducing $y'=\frac{\textrm{d}y}{\textrm{d}z}=\frac{p_y}{p_z}$,
where $p_y$ is the vertical component of the momentum, and noting
that $y'$ is a constant in a drift space for an
angular-momentum-dominated beam, we see that the change in vertical
coordinate is $\Delta y =y'D =r_2 \textrm{sin}\theta$ (see
Fig.~\ref{fig:camdrif}). Hence Eq.~(\ref{e:demomechL1}) can be
rewritten in the convenient form
\begin{eqnarray}\label{e:demomechL2}
\boldsymbol{L}=r_1p_z y'
\hat{\boldsymbol{e}}_z=p_z\frac{r_1r_2\textrm{sin}\theta}{D}\hat{\boldsymbol{e}_z}.
\end{eqnarray}
For a cylindrically symmetric laminar beam with rms transverse beam
sizes $\sigma_1$ and $\sigma_2$ at respective locations $z_1$ and
$z_2$ along the beamline, the averaged mechanical angular momentum
can then be calculated via
\begin{equation}\label{e:mechLsigma}
\mean{L} =  2p_z \frac{\sigma_1 \sigma_2 \sin{\theta}}{D}.
\end{equation}
Thus the measurements of rms beam sizes at locations $z_1$ and $z_2$
along with the corresponding shearing angle provide all the
necessary information for calculating the mechanical angular
momentum. Experimentally, the shearing angle is obtained by
inserting at location $z_1$ a multislit mask and measuring the
corresponding shearing angle of the beamlets at the location $z_2$;
see Fig.~\ref{fig:spotslit}. For the mechanical angular momentum
measurement reported here we use the diagnostic stations X3 and X6
(see Fig.~\ref{fig:injector}). The X3 diagnostic station includes an
OTR screen and two insertable multislit masks (with vertical and
horizontal slits). The station X6 is only equipped with an OTR
screen.

A set of measurements of mechanical angular momentum versus $B_0$
was reported in Ref.~\cite{YESPAC2003}. In the present Paper, such
measurements are performed by varying $B_0$ over a wider range ($B_0
\in [200,1000]$ Gauss for a bunch charge of $0.41 \pm 0.05$nC; see
details in Ref.~\cite{yesBD1}). The measurement technique discussed
in the previous paragraph was also numerically tested for each
experimental data point. In Fig.~\ref{fig:LvsLb} we compare the
measured mechanical angular momentum from Eq.~(\ref{e:mechLsigma})
with the canonical angular momentum calculated from
Eq.~(\ref{e:camdef}), given the $B_0$. The measured values include
both experimental data and simulated values, i.e., values that have
been retrieved after simulating the measurement technique
numerically with the particle tracking program {\sc
Astra}~\cite{astra}. The uncertainties in the measurement of angular
momentum are obtained via error propagation from the direct
measurements of rms beam sizes and the ``shearing angle''.

Conservation of canonical angular momentum is demonstrated in
Fig.~\ref{fig:LvsZ}, where the angular momentum was measured at
different locations along the beamline. In these measurements all
quadrupoles are turned off so that the beam propagated in a drift
space.

The dependence of mechanical angular momentum on the charge was also
explored. In this experiment, the laser spot size was set to
$\sigma_c= 0.82$~mm, and the laser intensity was varied via a
wave-plate attenuator located in the UV laser path. The results,
shown in Fig.~\ref{fig:LvsQ}{(a)}, indicate the mechanical angular
momentum, for our set of operating parameters, is
charge-independent, confirming our assumption that the beam dynamics
is angular-momentum-dominated in the range explored here.

Finally the dependence of canonical angular momentum versus
$\sigma_c$ was investigated. The laser intensity was held constant
and $B_0$ was identical to the previous experiment ($B_0=962$~G).
The charge density in the bunch is therefore kept constant. The
measurements [see Fig.~\ref{fig:LvsQ}{(b)}] support the expected
quadratic dependence of the angular momentum on $\sigma_c$ indicated
in Eq.~(\ref{e:camdef}).

The measured dependencies of canonical angular momentum on the
different parameters are all in good agreement with theoretical
expectations. Such an agreement gives us some confidence on our
ability to control the angular momentum of the incoming beam
upstream of the RTFB section.
\section{removal of angular momentum and flat-beam generation}~\label{sec:REMOVAL}
To remove angular momentum, it is necessary to apply a torque on the
beam. A quadrupole can exert a net torque only on an incoming
asymmetric beam. Thus more than one quadrupole is needed to remove
the angular momentum of an cylindrically symmetric incoming beam. A
first quadrupole followed by a drift space will introduce asymmetry
in the $x$-$y$ space, while the other quadrupoles downstream are
properly tuned to apply a total net torque such that the angular
momentum is removed at the exit of the quadrupole section. For the
series of measurements and simulations presented in this section, a
set of three skew quadrupoles (S2, S3, S5 in
Fig.~\ref{fig:injector}) are used to remove the angular momentum and
generate a flat beam.
\begin{figure}[hbt]\centering
\includegraphics[width=0.5\textwidth]{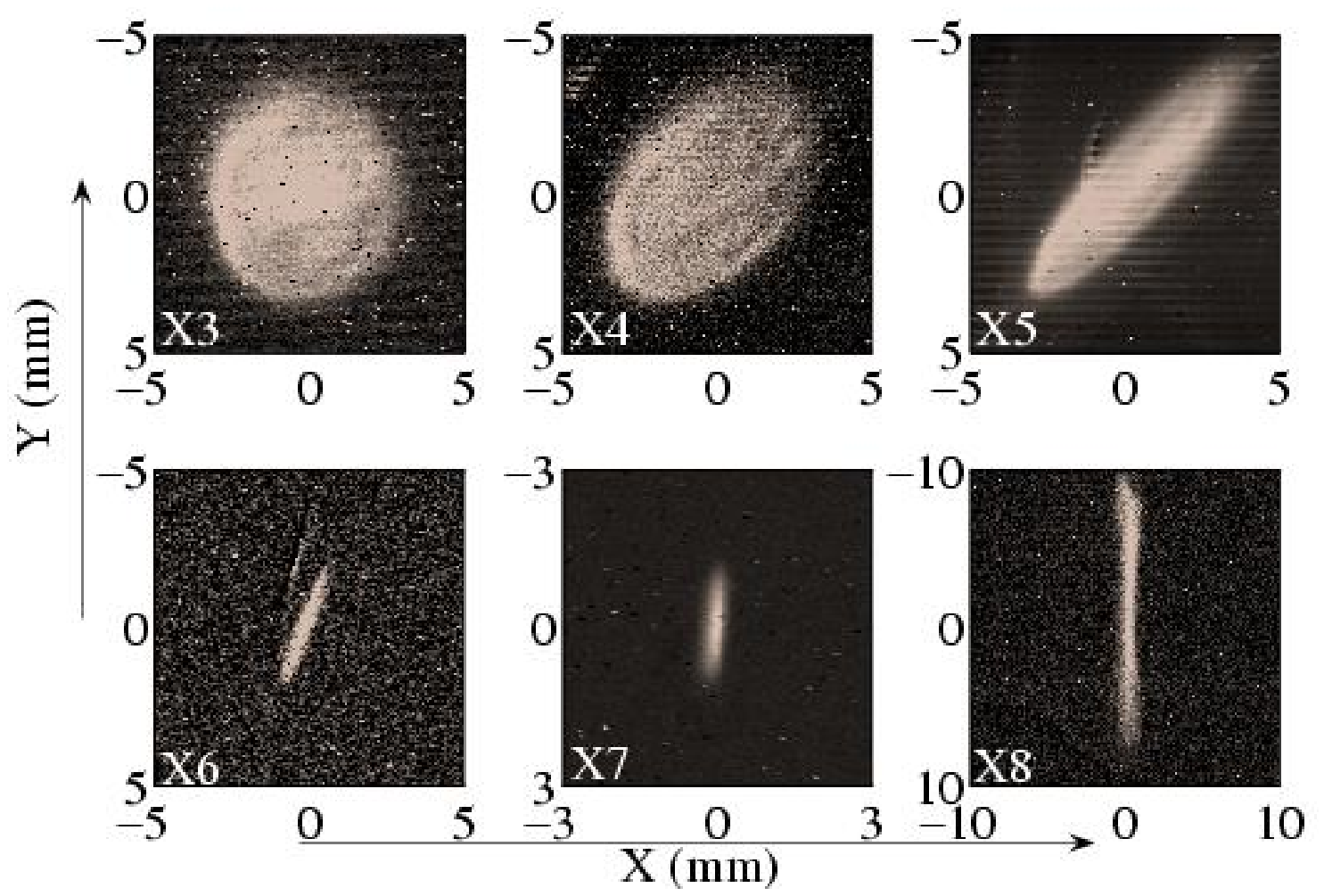}
\includegraphics[width=0.5\textwidth]{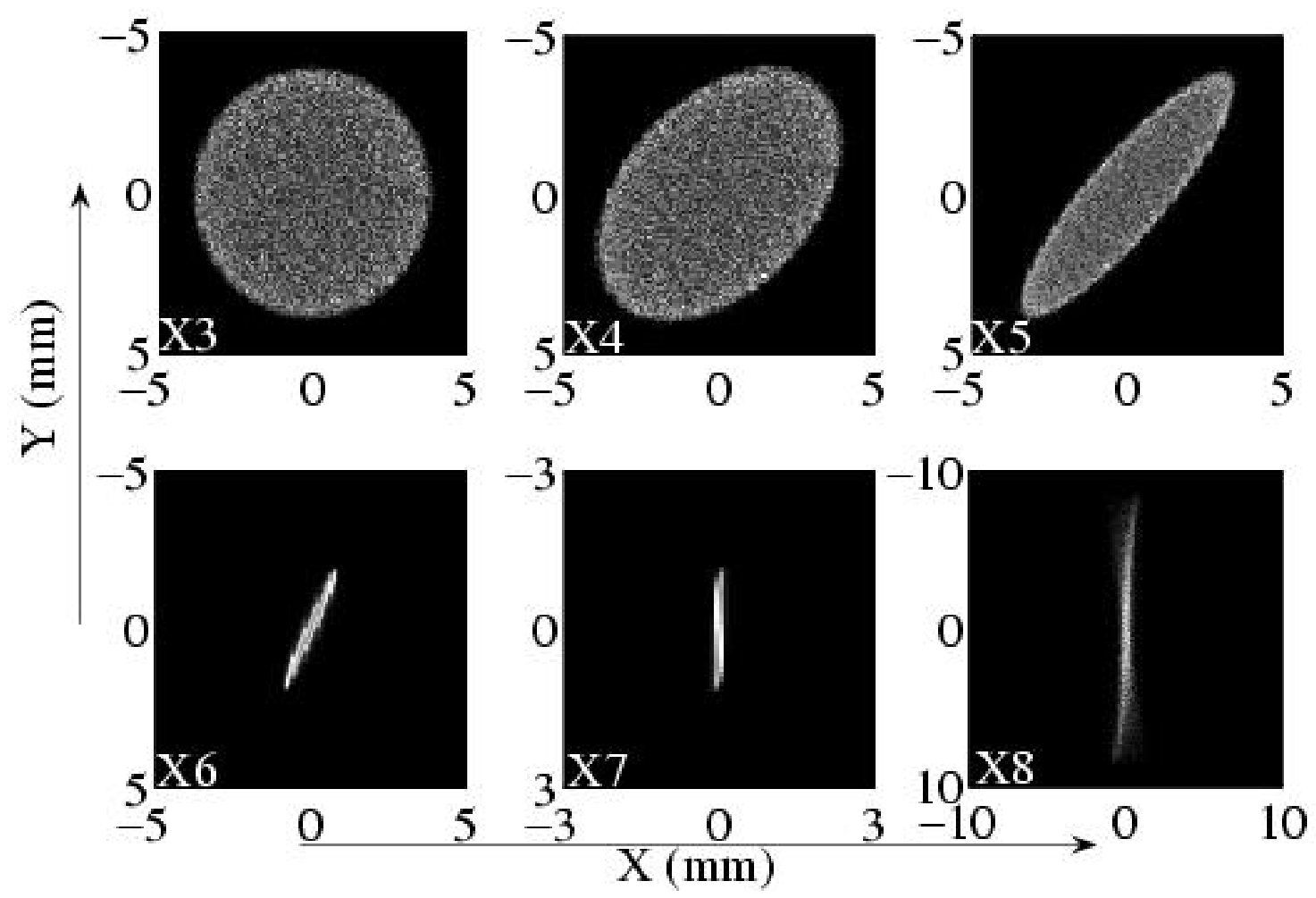}
\caption{Measured (top six photos) and simulated (bottom six plots)
beam transverse density evolution in the RTFB section. The
consecutive plots correspond to locations X3, X4, X5, X6, X7 and X8
shown in Fig.~\ref{fig:injector}. \label{fig:rftbsim}}
\end{figure}

Given the photoinjector parameters, numerical simulations of the
beamline (from the photocathode up to the entrance of the RTFB
transformer) are performed using {\sc Astra}. The four-dimensional
phase-space coordinates are propagated downstream of the transformer
using a linear transfer matrix. The initial values of the skew
quadrupole strengths are those derived, under the thin-lens
approximation, in Reference~\cite{flat2}. They are then optimized,
using a least-square technique, to minimize the $x$-$y$ coupling
terms of the beam matrix at the exit of the transformer. The final
optimized quadrupole strengths are used for subsequent \astra
simulation of the beam dynamics through the RTFB transformer.

Further empirical optimization around the predicted values is
generally needed to insure the angular momentum is totally removed,
as inferred by observation of the $x$-$y$ coupling at several
locations downstream of the RTFB section. Evolution of transverse
density throughout the RTFB section is in good agreement with
expectations from simulations, as shown in Fig.~\ref{fig:rftbsim}.
Each of the top six photos is a superposition of 5 bunches with
charge of 0.55 $\pm$ 0.10 nC. In the sequence of measurements and
simulations presented there, the incoming round beam (X3) is
transformed into a flat beam characterized by the large asymmetry
(X7 and X8). The mechanical angular momentum is removed: there is no
noticeable shearing as the beam propagates from X7 to X8.
\section{conclusion}~\label{sec:FUTURE}
We have experimentally explored some parametric dependencies of
angular momentum for an angular-momentum-dominated electron beam
produced in a photoinjector. The results obtained are in good
agreement with theoretical expectations, giving us some confidence
in our understanding of the angular-momentum-dominated beam.
\section{acknowledgements}
We wish to express our gratitude to H. Edwards for her many valuable
suggestions and stimulating discussions during the experiment, and
for her constant support. We are grateful to D. Edwards for his
comments on the manuscript and his leadership in the first flat beam
demonstration experiment. We are indebted to C. Bohn of Northern
Illinois University for carefully reading and commenting on the
manuscript.  We thank M. H\"uning, K. Desler for their help in the
operation, and W. Muranyi, M. Heinz, M. Rauchmiller, R. Padilla, P.
Prieto and B. Degraff for their excellent technical support. This
work was supported by Universities Research Association Inc. under
contract DE-AC02-76CH00300 with the U.S. Department of Energy, and
by NICADD.

\end{document}